\documentclass[traditabstract]{aa} 
\usepackage{graphicx}
\usepackage{txfonts}
\usepackage{wasysym}
\usepackage{array}
\usepackage{natbib}
\bibpunct{(}{)}{;}{a}{}{,} 
\usepackage{ulem}
\usepackage{aas}
\usepackage{bm}
\begin{document}
\title{The blue sky of GJ3470b: the atmosphere of a \\ low-mass planet
       unveiled by ground-based photometry\thanks{Based 
on data acquired using the Large Binocular Telescope
(LBT). The LBT is an international collaboration among institutions 
in the United States, Italy, and Germany. LBT Corporation 
partners are the University of Arizona on behalf of the Arizona 
university system; Istituto Nazionale di Astrofisica, Italy;
LBT Beteiligungsgesellschaft, Germany, representing the Max-Planck 
Society, the Astrophysical Institute Potsdam, and Heidelberg 
University; the Ohio State University; and the Research
Corporation, on behalf of the University of Notre Dame, University 
of Minnesota and University of Virginia.}
}
\author{V.~Nascimbeni\inst{1,2}\thanks{email 
        address: \texttt{valerio.nascimbeni@unipd.it}}
   \and G.~Piotto\inst{1,2} 
   \and I.~Pagano\inst{3} 
   \and G.~Scandariato\inst{4}  
   \and E.~Sani\inst{5}
   \and M.~Fumana\inst{6} 
       }
\institute{Dipartimento di Fisica e Astronomia, 
           Universit\`a degli Studi di Padova,
           Vicolo dell'Osservatorio 3, 35122 Padova, Italy
      \and
          INAF -- Osservatorio Astronomico di Padova, 
          vicolo dell'Osservatorio 5, 35122 Padova, Italy
      \and
          INAF -- Osservatorio Astrofisico di Catania, 
          via S. Sofia 78, 95123 Catania, Italy
      \and
          INAF -- Osservatorio Astronomico di Palermo, 
          Piazza del Parlamento 1, 90134 Palermo, Italy 
      \and
          INAF -- Osservatorio Astrofisico di Arcetri, 
          largo E.~Fermi 5,  50125, Firenze, Italy
      \and 
          INAF -- Istituto di Astrofisica Spaziale e Fisica Cosmica 
          Milano, via Bassini 15, 20133 Milano, Italy
          }

   \date{Submitted 2013 May 27; Accepted 2013 Aug 30.}

\abstract{GJ3470b is a rare example of a ``hot Uranus''  transiting
  exoplanet orbiting a nearby M1.5 dwarf. It is of crucial interest
  for atmospheric studies because it is one of  the most inflated
  low-mass planets known, bridging the boundary between
  ``super-Earths'' and Neptunian planets.  We present two new
  ground-based light curves of GJ3470b gathered by the LBC camera at
  the Large Binocular Telescope.  Simultaneous photometry in the
  ultraviolet ($\lambda_\mathrm{c}=357.5$ nm)  and optical infrared
  ($\lambda_\mathrm{c}=963.5$ nm) allowed us to detect a significant
  change of the effective radius of GJ3470b as a function of
  wavelength. This can be interpreted as a signature of scattering
  processes occurring in the planetary atmosphere, which should be
  cloud-free and with a low mean molecular weight.  The unprecedented
  accuracy of our measurements demonstrates that  
  the photometric detection of Earth-sized planets around M dwarfs
  is achievable using 8-10m size ground-based telescopes.
  We provide updated planetary parameters, and a greatly
  improved orbital  ephemeris for any forthcoming study of this
  planet.}

   \authorrunning{Nascimbeni et al.}
   \titlerunning{The blue sky of GJ3470b}
   \maketitle


\newcommand{\fo}{f_\mathrm{o}}
\newcommand{\fu}{f_\mathrm{u}}
\newcommand{\ic}{I_\mathrm{c}}
\newcommand{\rc}{R_\mathrm{c}}
\newcommand{\uspec}{U_\mathrm{spec}}
\newcommand{\lambdac}{\lambda_\mathrm{c}}
\newcommand{\mplanet}{M_\mathrm{p}}
\newcommand{\mearth}{M_\oplus}
\newcommand{\rearth}{R_\oplus}
\newcommand{\msun}{M_\odot}
\newcommand{\rsun}{R_\odot}
\newcommand{\mstar}{M_\star}
\newcommand{\rstar}{R_\star}
\newcommand{\mjup}{M_\mathrm{jup}}
\newcommand{\rjup}{R_\mathrm{jup}}
\newcommand{\rplanet}{R_\mathrm{p}}
\newcommand{\rplanetzero}{R_\mathrm{p,0}}
\newcommand{\rhoplanet}{\rho_\mathrm{p}}
\newcommand{\de}{\mathrm{d}}
\newcommand{\teq}{T_\mathrm{eq}}
\newcommand{\kb}{k_\mathrm{b}}
\newcommand{\htwo}{\mathrm{H}_2}
\newcommand{\htwoo}{\mathrm{H}_2\mathrm{O}}
\newcommand{\chfour}{\mathrm{CH}_4}
\newcommand{\der}{\de\rplanet/\de\ln\!\lambda }
\newcommand{\chisq}{\chi^2}
\newcommand{\chisqr}{\chi^2_\mathrm{r}}
\newcommand{\teff}{T_\mathrm{eff}}
\newcommand{\logg}{\log g}
\newcommand{\feh}{[\mathrm{Fe}/\mathrm{H}]}

\section{Introduction}\label{introduction}

The number of discovered exoplanets is growing at a very fast pace,
reaching more than seven-hundred in May 2013\footnote{according to
  \texttt{exoplanets.org} \citep{exoplanets.org}.}. Nearly one-third
of them transit in front of their host stars, a lucky circumstance
that allows us to access their radius through simple geometrical
assumptions.  Nevertheless, the great majority of these planets lack a
complete characterization, which should include the knowledge of their
internal structure, atmospheric composition and evolutionary history
\citep{seager2010}. This is even more limiting if one takes into
account that most extrasolar planets known so far have physical
properties completely different from  those found in our Solar System,
i.~e. we cannot rely on ``easy''  analogies \citep{zhou2012}.

Most discoveries of the early years (1995-2007) were strongly biased
towards ``hot Jupiter''-type planets, i.e.~gas giants orbiting at a
few stellar radii from their host star. Thanks to  dedicated space
missions such as CoRoT \citep{baglin2006} and Kepler
\citep{borucki2010}  we are now accessing a broader region of the
parameter space, probing smaller  planetary radii and masses, and
detecting planetary systems hosted by a wider variety of parent
stars. Not surprisingly, the observational landscape is getting even
more complicated, with wholly new --and sometimes unexpected-- classes
of planets  appearing.

``Super-Earths'' ($2\lesssim\mplanet\lesssim 10$ $\mearth$) and
``Neptunian'' planets ($15\lesssim\mplanet\lesssim 50$ $\mearth$) are
now puzzling theoreticians, revealing a complex diversity which
cannot be easily explained by models or constrained by observations
\citep{hagh2011}.  Super-Earths and Neptunes were once hypothesized to
be separate and  well defined classes of rocky and icy planets,
respectively. Now we realize that their measured densities range from
$\rho_\mathrm{p}=0.27$ (Kepler-18d; \citealt{cochran2011}) 
to 
$\sim 10$~$\mathrm{g}\,\mathrm{cm}^{-3}$ (Kepler-10b; \citealt{batalha2011}),
with a significant overlap between
super-Earths and Neptunes \citep{exoplanets.org}. 
In this mass range, average  densities
are unable to put a firm constraint on the inner structure of the
planet, and theoretical models can lead to several  degenerate
solutions (e.g., \citealt{adams2008}). One of the missing key quantities is
the atmospheric scale height $H$ of the planet, defined, 
for a well-mixed isothermal atmosphere in equilibrium, as 
\begin{equation}
H=\frac{\kb\teq}{\mu g} \textrm{ ,}
\label{scaleh}
\end{equation}
where $\kb$ is the Boltzmann constant, 
$\teq$ the equilibrium temperature at its surface, $\mu$ the mean
molecular weight of the atmosphere, and $g$ the planet surface gravity. 

An emblematic case is that of GJ1214b
\citep{charbonneau2009}, for which allowed scenarios include a
``mini-Jupiter'' with a small solid core and large envelopes of
primordial H/He, a Neptune-like planet with an atmosphere made of
sublimated ices, a ``water world'', or a rocky super-Earth with an
outgassed atmosphere \citep{rogers2010}. Newer examples of such
degeneracy are HAT-P-26b \citep{hartman2011} and GJ3470b \citep{bonfils2012}.
Additional information other
than $\rhoplanet$  is required to break the model
degeneracy, and an effective way to do it is to search for
spectral signatures coming from the planetary atmosphere
to probe its composition, or to put constraints on $\mu$
\citep{seager2010}.  

Additional information can come from \emph{transmission spectroscopy}, 
a technique based on the observation of
exoplanetary transits at different wavelengths, searching
for changes of its effective radius  $\rplanet$ as a function
of $\lambda$ due to absorption and/or scattering processes undergone
by stellar light after traveling through the atmospheric limb \citep{sing2011}.
Of course, low-resolution transmission spectroscopy  can also be performed
at a higher efficiency by gathering photometric  light curves through
different passbands and investigating the resulting 
$\rplanet(\lambda)$ dependence (e.g., \citealt{demooij2012}).

While most frequently transmission spectroscopy is carried out to
search for atomic or molecular absorption features, useful 
informations can be also inferred  from the detection of
scattering processes affecting the continuum. The most straightforward
one to interpret is Rayleigh scattering, due to molecules such as
$\htwo$ \citep{lecavelier2008b} or  to haze from condensate particles
\citep{fortney2005,lecavelier2008a}.  
Rayleigh scattering has a typical signature: its steep  
$\lambda^{-4}$ dependence, which translate into a larger
$\rplanet$ on the blue side of the optical spectrum.
Under simplified assumptions,
we can analytically approximate the 
scale height $H$ (and hence $\mu$, once a $\teq$ is assumed) 
from the steepness of $\rplanet (\lambda)$ where Rayleigh scattering
is dominant (\citealt{lecavelier2008a}, among others).  On the other hand, 
detailed atmospheric models are available, at the price of
increased complexity and number of free parameters
\citep{howe2012}.

GJ3470b has been defined a ``hot Uranus'' planet by its
discoverers, who first detected the 3.33-day radial velocity (RV) 
modulation of its M1.5V host star with HARPS and then caught its 
transits through ground-based photometry
\citep{bonfils2012}. Very recently, a follow-up analysis 
based on Spitzer, WIYN-3.5m and Magellan data  \citep{demory2013} resulted in 
updated values for the mass and radius of GJ3470b: 
$\mplanet = 13.9\pm 1.5$~$\mearth$ and 
$\rplanet = 4.83\pm 0.21$~$\rearth$, making it one of the less
dense low-mass planets known with 
$\rhoplanet = 0.72\pm 0.13$~$\mathrm{g}\,\mathrm{cm}^{-3}$.
Showing this peculiarity, having a total mass at the boundary between
super-Earths and Neptunes, GJ3470b stands out as an ideal case 
to test the present planetary structure and evolution theories, 
which predict for it a significantly extended envelope of 
primordial H and He \citep{demory2013,rogers2010} and thus a large scale
height. From the observational point of view, GJ3470b is 
promising thanks to its relatively large transit depth
$(\rplanet/\rstar)^2 \simeq 0.006$, 
being hosted by a small star ($\mstar\simeq 0.54$~$\msun$, 
$\rstar\simeq 0.57$~$\rsun$; \citealt{demory2013}). Yet the
level of chromospheric activity of the GJ3470 
appears to be low given its spectral type, showing less than 0.01
mag long-term photometric variability in the $\ic$ band \citep{fukui2013}.

In this paper we present new photometric data from LBT,
consisting of two simultaneous transit light curves of
GJ3470b gathered through passbands at the extremes of the optical spectrum 
($\lambdac = 357.5$ and 963.5 nm). Our primary aim was 
to search for signatures of Rayleigh scattering in the planetary
atmosphere. We describe the instrumental setup and observing strategy
in Sect.~\ref{observations}, and our data reduction techniques in 
Sect.~\ref{reduction}. Next we illustrate how we extracted the
photometric parameters of GJ3470b from our light curves
(Sect.~\ref{analysis}) including a correction on $k=\rplanet/\rstar$ 
due to the presence of unocculted spots. 
Finally, we discuss our results in Sect.~\ref{discussion}.

\section{Observations}\label{observations}

We observed a full transit of GJ3470b at LBT 
(Mt.~Graham, Arizona) in the night between 2013 Feb 16 and 17,
under a clear, photometric sky.
The Large Binocular Camera (LBC; \citealt{giallongo2008}) 
consists of two prime-focus, wide-field imagers mounted 
on the left and right arm of LBT, and optimized for blue
and red optical wavelengths, respectively.  
We exploited their full power to gather two simultaneous 
photometric series, setting the $\uspec$ filter on the blue 
channel (a filter with Sloan $u$ response but having an 
increased efficiency, centered at $\lambdac=357.5$ nm) and
the $F972N20$ filter in the red channel (an intermediate-band filter
centered at $\lambdac=963.5$ nm, originally designed for 
cosmological studies)\footnote{The FHWM of the 
$F972N20$ band is 22 nm; the passbands of both filters are 
tabulated in \texttt{http://abell.as.arizona.edu/\textasciitilde{}lbtsci/ Instruments/LBC/lbc\_description.html\#filters.html}
and plotted in the bottom panels of Fig.~\ref{spectrum} and \ref{slope}.}. 
These choice of passbands have been made to maximize
the wavelength span of our observations while at the same time
avoiding most of telluric lines in the red channel.

We set a constant exposure time of 60 s on both channels. 
We chose to read out only a $2300\times1800$ pixel
window from the central chip of each camera (\texttt{chip\#2}) 
in order to decrease the technical overheads and optimize the
efficiency of the series, achieving a 60\% duty-cycle 
on average. The resulting $7.6'\times 6.7'$ field of view (FOV) 
was previously tailored to include a suitable set of reference stars, 
required to perform differential photometry with a S/N
as high as possible in both channels. 
Stellar PSFs were defocused to $\sim 7''$ FWHM (full width at half maximum; 31 physical pixels)
in $\uspec$ and  $\sim 13''$ (58 pix) in $F972N20$ to avoid saturation and 
to minimize intra-pixel and pixel-to-pixel systematic 
errors \citep{nascimbeni2011a}. Autoguiding is not 
allowed on LBC during a photometric series; instead,
the telescope was passively tracking throughout the series, drifting
by about 80 pixels from the first to the last frame.

A total of 134 $\uspec$ frames and 146 $F972N20$
frames were secured from 3:22 to 7:06 UT, covering also 55+55 minutes 
of off-eclipse photometry before and after the external contacts
of the transits. 

\section{Data reduction}\label{reduction}

Both sets of frames were bias-corrected and flat-fielded using
standard procedures and twilight master flats. Then we extracted
the $\uspec$ and $F972N20$ light curves of GJ3470
using STARSKY. 

\begin{figure*}
\centering
\includegraphics[width=\columnwidth]{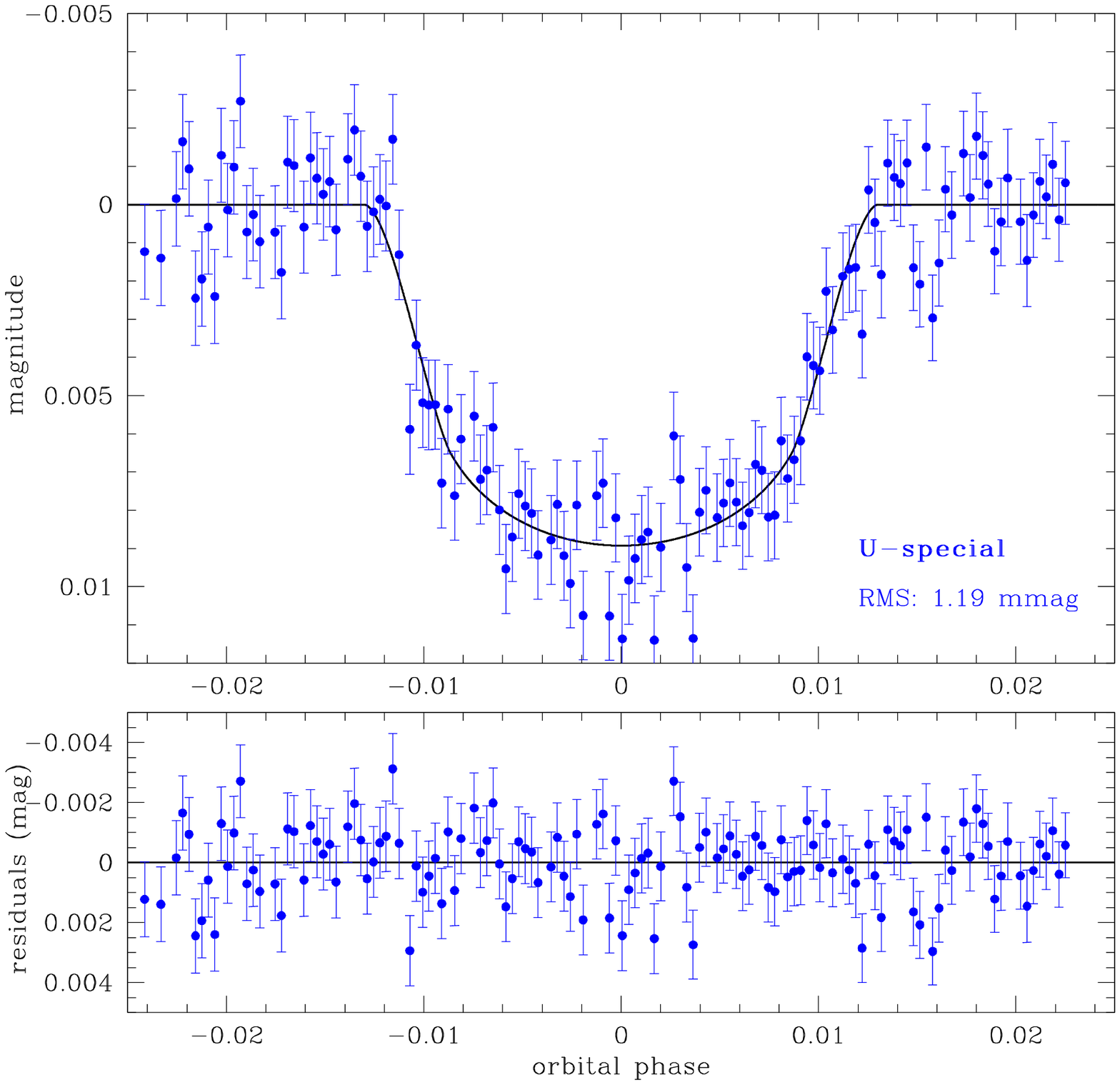}
\includegraphics[width=\columnwidth]{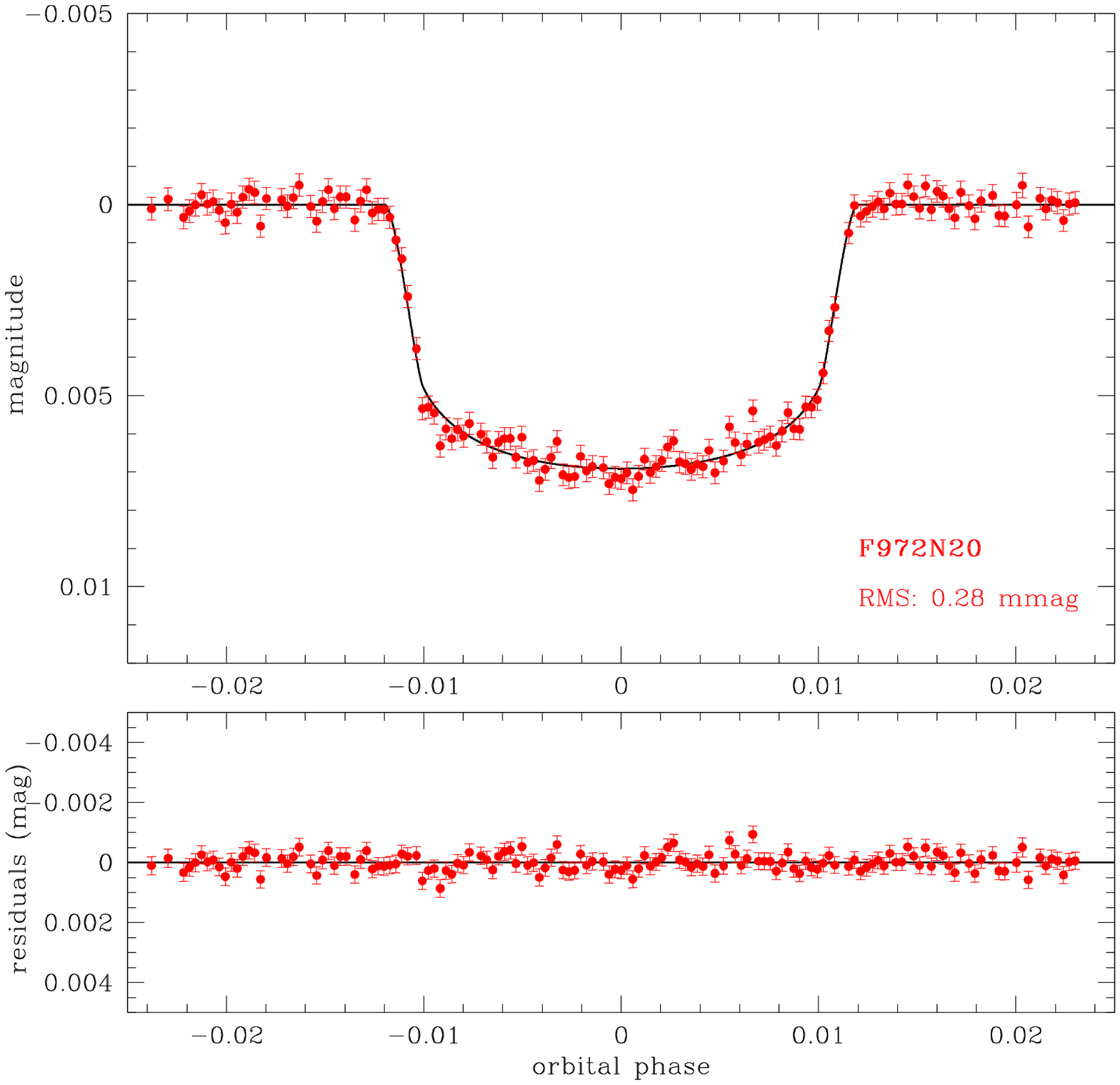}
\caption{\emph{Left plot:} light curve of GJ3470b 
gathered with the blue channel of LBC in the $\uspec$ band, 
plotted with the original sampling cadence (upper panel; the black
line corresponds to the best-fit model adopted in Table \ref{results})
and residuals from the best-fit model (lower 
panel). \emph{Right plot:} 
same as the left plot, but in the $F972N20$ band of the red channel
of LBC. The two light curves are simultaneous.}
\label{lcs}
\end{figure*}

STARSKY is a photometric pipeline optimized 
to perform  high-precision differential aperture photometry over 
defocused images, originally developed 
for the TASTE project (The Asiago Search for 
Transit timing variations of Exoplanets; \citealt{nascimbeni2011a,nascimbeni2011b}). 
Its prominent feature is an empirical algorithm to choose the 
aperture radius and to weight the flux from the comparison stars in 
such a way that the off-transit scatter of the target star is 
minimized \citep{nascimbeni2013a}. In a second step, STARSKY searches
for linear correlations between the differential magnitude and a set of
standard external parameters, including the $(x,y)$ position of
the star, its FWHM, sky background and time $t$. In our case, a
22.8-pixel ($5.1''$) aperture was selected as the optimal one 
for the $\uspec$ data set, followed by a decorrelation against FWHM 
and $t$. For the $F972N20$ series, the optimal aperture diameter 
was 40.0 pixels ($9''$), followed by a decorrelation against $t$. 
It is remarkable that we did not detect any significant $(x,y)$-dependent 
systematic effect neither in the blue channel nor in the red one,
suggesting that the ``extreme'' defocusing strategy we applied is
effective against most residual flat-field and intrapixel errors.

The resulting light curves are plotted in Fig.~\ref{lcs} in their
original cadence. Their overall root mean square (RMS) scatter is 1.19 mmag ($\uspec$)
and 0.28 mmag ($F972N20$) over a mean cadence of 95 s, only slightly
larger than the theoretical values estimated by standard ``white
noise'' formulae: 1.08 mmag and 0.24 mmag, respectively
\citep{howell2006}.  It is worth noting that the $F972N20$ series is,
to our knowledge,  the most precise light curve ever obtained from a
ground-based facility, breaking by a narrow margin the  previous
record set by \citet{tregloan2013} at the New Technology Telescope
(NTT).  The prospects from these
achievements will be further  discussed in Sect.~\ref{discussion}.

\section{Data analysis}\label{analysis}

\subsection{Light curve modeling}

The JKTEBOP
code\footnote{\texttt{http://www.astro.keele.ac.uk/\textasciitilde{}jkt/codes/jktebop.html}}
version 28 \citep{southworth2004} was employed to fit a transit model
over the LBC light curves. We chose to derive the uncertainties over
the best-fit parameters with a Monte Carlo bootstrap algorithm
\citep{southworth2005}, to avoid underestimation due to correlations
between parameters and  residual systematic noise (also called ``red
noise''; \citealt{pont2007}).  A quadratic law for modeling the
stellar limb darkening (LD) is adopted \citep{claret2004}. 

We first fitted the $F972N20$ curve leaving five free parameters: the
radius  ratio $k=\rstar/\rplanet$, the sum of the planet and star
radii scaled by the orbital semi-major axis
$\Sigma=(\rstar+\rplanet)/a$, the orbital inclination  $i$, the
central time $T_0$ and the linear LD coefficient $u_1$. Following
a common practice to improve the robustness of
the fit \citep{southworth2008}, we did not fit for the quadratic LD coefficient $u_2$,
because it is strongly degenerate with $u_1$. 
Even fitting for the weakly correlated parameters
$c_1 = 2u_1 + u_2$ and $c_2 = u_1 - 2u_2$ 
(as suggested by \citealt{holman2006}) 
or other linear combinations of $u_1$ and $u_2$ 
leads to unphysical results due to overfit.
Instead, $u_2$ was injected in the Monte Carlo bootstrap as a Gaussian prior,
extracting its theoretical value from the tables by \citet{claret2012}, 
interpolated by adopting the stellar atmospheric parameters published by
\citet{demory2013} and taking into account their 
uncertainties on $\teff$, $\logg$ and $\feh$.
As the $F972N20$ band is 
not tabulated by \citet{claret2012}, we adopted the SDSS $z'$ values
basing on the fact that $z'$ and $F972N20$ overlap with each other in
a spectral region where the specific intensity $I_\nu(\theta)$ 
is nearly devoid of spectral features
($\theta$ being the angle between the observer and the normal 
to the stellar surface). Moreover, for our given set of atmospheric
parameters,
the quadratic term $u_2$ is nearly insensitive to $\lambda$, 
oscillating between 0.260 and 0.366 over the full optical-NIR range between 
Sloan $u$ to 2MASS $J$, with a mean and RMS of $0.30\pm0.03$. 
After a first best-fit solution is found on
real data, we ran a bootstrap analysis on $20\,000$ resampled light
curves having the same statistical properties of the real one;  the
final best-fit values and the asymmetric errors on each fitted
quantity  were then estimated from the 50$^\mathrm{th}$,
15.87$^\mathrm{th}$, and $84.13^\mathrm{th}$  percentile of the
bootstrapped distribution (Table \ref{results}, second column).  We
emphasize that the fitted value of $u_1=0.25\pm 0.04$ is perfectly
consistent  with the theoretical estimate $u_{1,\mathrm{th}}=0.275$ 
for the $z'$ band by
\citet{claret2012}, supporting our previous assumption 
$u_2 (z') \simeq u_2 (\mathrm{F972N20})$. As a 
further cross-check, we verified that even by setting the input $u_2$ 
at the aforementioned extremal values 0.26 and 0.37, the resulting
best-fit parameters change by less than 1-$\sigma$ with respect to our
adopted solution.

The $\uspec$ light curve has a much lower $S/N$ compared to the 
$F972N20$ one. We then improved the robustness of the fit 
by fixing the value of 
$i$ (which is a purely geometrical parameter, not dependent on $\lambda$) 
to that found on the $F972N20$ series ($88.1^\circ\pm 0.3$), 
by injecting a Gaussian distribution with the same mean and standard 
deviation into the bootstrap analysis. The same is done for $T_0$.
Instead, $\Sigma$ (and so derived quantities such as $a/\rstar$)
was left free, since a small dependence of these
variables on $\lambda$ is expected when $\rplanet (\lambda)$ itself is
not constant.
The value of both $u_1$ and $u_2$ are fixed to their \citet{claret2012} $u'$ 
theoretical Gaussian priors as above. The resulting best-fit values for the two fitted 
parameters ($\Sigma$ and $k$) are reported in the third column of Table
\ref{results}. Other quantities of interests, such as the impact parameter $b$
and the full duration $T_{14}$ \citep{winn2009} are extracted 
for both LBC light curves from the  
bootstrapped distribution, assuming zero eccentricity ($e=0$) as
suggested by the RV measurements \citep{bonfils2012}.

We note that all the best-fit parameters we estimated on both
LBC curves are in excellent agreement with each other and 
with the \citet{demory2013} values (Table~\ref{results})
within their errorbars. The only significant exception is $k$ 
which is offset respectively by
\begin{equation}
\frac{k(\uspec)-k(4.5\mu\mathrm{m})}{\Delta k(\uspec)} \simeq 3.6\,\sigma \textrm{ ,}\quad
\frac{k(F972)-k(4.5\mu\mathrm{m})}{\Delta k(F972 )} \simeq -6.4\,\sigma 
\end{equation}
This offset is not driven by a different assumption on the 
orbital inclination (or, equivalently, on the impact parameter), 
as our best-fit value $i=88^\circ.1\pm0.3$
is in perfect agreement with $i=88^\circ.3\pm 0.5$ found by \citet{demory2013} and 
later adopted by \citet{fukui2013} as a prior in their analysis. As a further
check, we reran our analysis adopting a Gaussian prior $i=88^\circ.3\pm 0.5$, obtaining
statistically indistinguishable values of $k$ for both light curves 
($\Delta k/\sigma(k) \leq 0.3$).
As suggested by the anonymous referee, we also redid a full analysis 
on both light curves by injecting Gaussian priors on $c_1$ and $c_2$ defined as above, and leaving everything else unchanged. 
Again, the resulting best-fit parameters and their associated errors are fully consistent with those listed in Table~\ref{results}, meaning that correlation
between the LD parameters cannot explain the measured discrepancy between $k(\uspec)$
and $k(F972)$.
In the next subsection we will apply a correction to our
$k$ values to take into account the presence of unocculted
starspots on the GJ3470 photosphere. Then in Sect.~\ref{discussion}
we will discuss the significance and origin of this chromatic signature.

\begin{figure*}
\centering
\includegraphics[width=1.7\columnwidth]{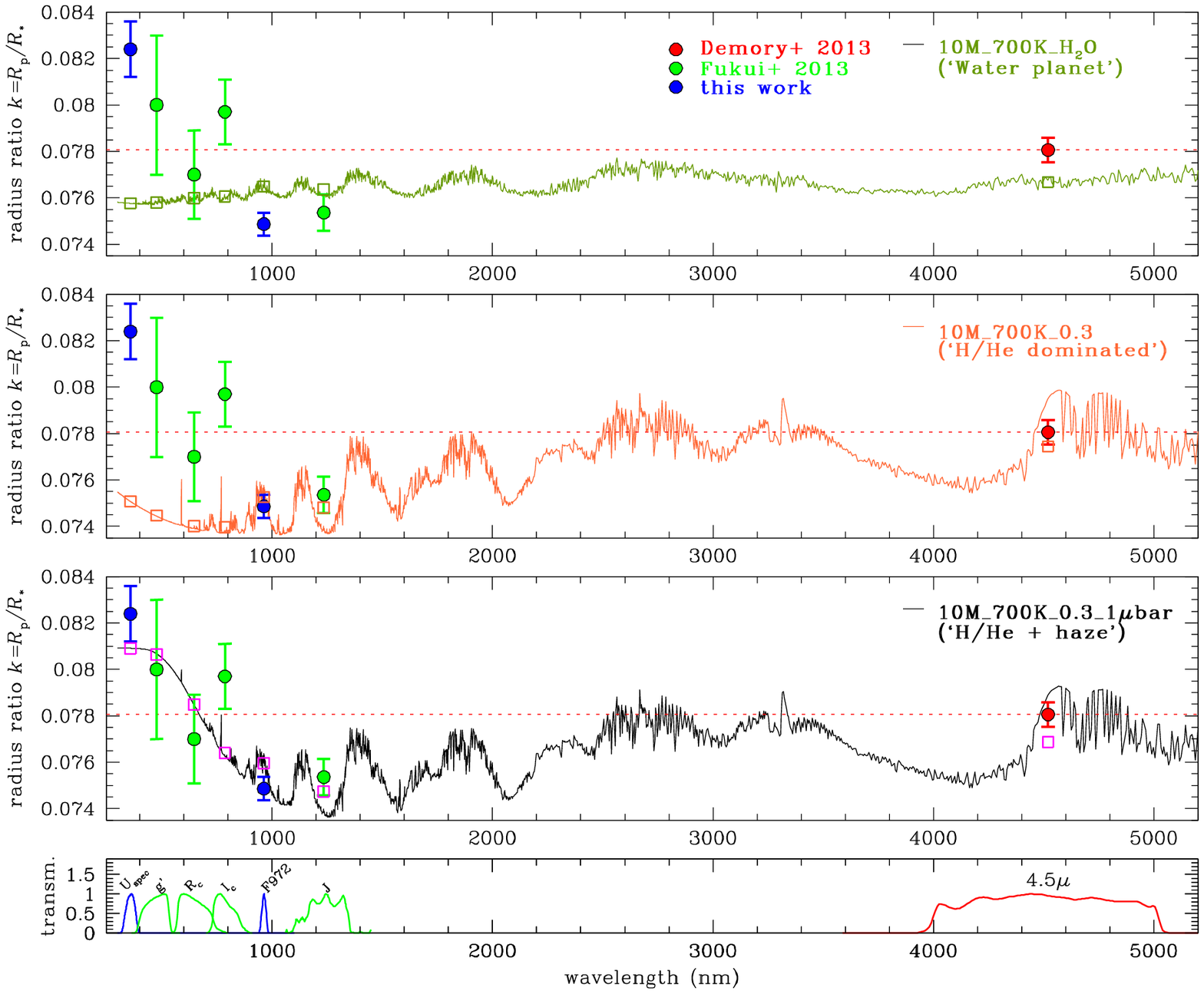}
\caption{\emph{Upper panel:} Reconstructed transmission spectrum of
  GJ3470b. The $U_\mathrm{spec}$ and $F972N20$ data points are
  extracted from our LBC light curves (blue circles), the
  $g'R_\mathrm{c}I_\mathrm{c}J$ points from \citet{fukui2013} (green
  circles), and the $4.5\mu$m point from \citet{demory2013} (red
  circle). 
  The model plotted with a dark green line is scaled from a
  \citet{howe2012} model computed for a 10-$M_\oplus$ cloud-free,
  haze-free planet with $T_\mathrm{eq}=700$ K and a pure water ($\htwoo$)
  atmosphere.
  The open squares represent the model points integrated over 
  the instrumental passbands. 
  \emph{Second panel:}  same as above, but with a
  cloud-free, haze-free, metal-poor atmosphere 
  dominated by H and He, computed by the same authors (orange line).
  Abundances are scaled from the solar ones 
  ($Z = 0.3\times Z_\odot$)
  \emph{Third panel:} same as above, but adding the contribution of
  a scattering haze from Tholin particles  at 1 $\mu$bar. 
  \emph{Lower panel:}
  Instrumental passbands employed for each data point.}
\label{spectrum}
\end{figure*}

The extremely high precision on $T_0$ we achieved on the $F972N20$ 
transit (about 9 s) allowed us to compute a greatly improved orbital
ephemeris of GJ3470b, by combining our measurement with the two $T_0$
published by \citet{demory2013} and fitting a linear ephemeris by ordinary weighted
least squares:
\begin{equation}
\begin{array}{rcl}
T_0 (\mathrm{BJD}_{\mathrm{TDB}}) &= &(2\,456\,340.72559 \pm 0.00010 ) + \\
               & + &(3.336\,649\pm 0.000\,002)\cdot N \\
\end{array}
\end{equation}
where the integer epoch $N$ is the number of transits elapsed from our
LBC observation. The combination of time standard and reference frame here
adopted is $\mathrm{BJD}_{\mathrm{TDB}}$ (Barycentric Julian Day, 
Barycentric Dynamical Time), as we follow the prescription
by \citet{eastman2010}.

\begin{table*}
\caption{Summary of the best-fit parameters found for GJ3470b.}
\centering\setlength{\extrarowheight}{3pt}
\begin{tabular}{l|c|cc|c} \hline \hline
parameter                                        & \citet{demory2013} & this work & this work & unit \\ 
                                                 & (IRAC $4.5\mu$)                 & (LBC $F972N20$)                 & (LBC $U_\mathrm{spec}$) & \\ \hline
central wavelength $\lambda_\mathrm{c}$          & 4520                            & 963.5                           & 357.5 & nm \\       
radius ratio $k=R_\mathrm{p}/R_\star$            & $0.07806^{+0.00052}_{-0.00054}$ & $0.07484^{+0.00052}_{-0.00048}$ & $0.0821\pm 0.0013$ & --- \\
sum of radii $\Sigma = (R_\mathrm{p}+R_\star)/a$ &                                 & $0.0819\pm 0.0025$              & $0.0822^{+0.0031}_{-0.0029}$ & --- \\
 $a/R_\star$                                     &                                 & $13.13^{+0.42}_{-0.38}$         & $13.18\pm 0.47$ & --- \\
orbital inclination $i$                          & $88.3^{+0.5}_{-0.4}$            & \multicolumn{2}{c|}{$88.12^{+0.34}_{-0.30}$ (a)} & degrees \\
impact parameter $b=a\cos i / R_\star$           & $0.40^{+0.06}_{-0.08}$          & $0.43^{+0.05}_{-0.07}$          & $0.44\pm 0.06$ & --- \\
eccentricity $e$                                 & $0.0^{\mathrm{(b)}}$            & $0.0^{\mathrm{(b)}}$            & $0.0^{\mathrm{(b)}}$ & --- \\
linear LD coeff.~$u_1$                           & $0.013\pm 0.015$                & $0.25\pm 0.04$                  & $0.49\pm 0.03^{\mathrm{(c)}}$ & --- \\
quadratic LD coeff.~$u_2$                        & $0.181\pm 0.010$                & $0.30\pm 0.03^{\mathrm{(c)}}$   & $0.29\pm 0.03^{\mathrm{(c)}}$ & --- \\
transit duration $W=T_{14}$                      & $0.0790\pm 0.0005$              & $0.0798^{+0.0004}_{-0.0005}$    & $0.0799^{+0.0016}_{-0.0014}$ & days \\
                                                 & $113.8\pm 0.7$                  & $114.9^{+0.5}_{-0.7}$           & $115.0^{+2.3}_{-2.1}$ & minutes \\
central time $T_0$                               & 6090.47690 $\pm$ 0.00015        & \multicolumn{2}{c|}{6340.72559 $^{+0.00009}_{-0.00011}$ (a)} & BJD(TDB)$^{\mathrm{(d)}}$ \\
orbital period $P$                               & $3.33671\pm 0.00005$            & \multicolumn{2}{c|}{$3.336\,649\pm 0.000\,002^{\mathrm{(e)}}$} & days \\ \hline
\end{tabular}
\tablefoot{\emph{(a)}: fitted on the $F972N20$ light curve and then assumed as input distribution for the $\uspec$ Monte Carlo
simulation (see text for details). \emph{(b)} fixed to zero on all light curves. \emph{(c)} fixed as Gaussian priors to the theoretical values interpolated
from \citet{claret2012}, assuming the atmospheric parameters estimated by \citet{demory2013} and their uncertainties. \emph{(d)} BJD(TDB) $-$ $2\,450\,000$ \citep{eastman2010}.
\emph{(e)} linear ephemeris evaluated by combining our $T_0$ with that published by \citet{demory2013}.}
\label{results}

\end{table*}


\subsection{Correction for unocculted starspots}

 Starspots affect in two ways the stellar
 flux variations in a transit light curve. Spots that are not occulted
 by the planet produce a de\-crea\-se in the out-of-transit flux of
 the star, which is the reference level to nor\-ma\-li\-ze the transit
 profile. Spots that are occulted during the transit produce a
 relative increase in the stellar flux because the flux blocked by the
 planetary disc is lower than in the case of the unperturbed
 photosphere.  Therefore, if the star is spotted,  and $\fo$ and $\fu$
 are the spot filling factors of the occulted and unocculted
 photosphere, respectively,   the relative radius of the planet $k_a$
 derived from the fitting of the transit profile, is related to the
 true relative radius $k$ by:
 \begin{equation}
 k_a \simeq  k \left[ 1  - \frac{1}{2} A_\lambda (\fo - \fu)
   \right] \textrm{ ,}
 \label{ratioff} \end{equation}
 where $A_{\lambda} \equiv 1-F_\mathrm{s}(\lambda)/F_\star(\lambda)$, 
with $F_\star(\lambda)$ and  $F_\textrm{s}(\lambda)$ 
the emerging fluxes from the unperturbed
 photosphere, and  the spotted area considered as a photospheric
 model computed at the mean spot temperature,  respectively.
 
 \citet{bonfils2012} suggest that  GJ3470 is not a very active star
 according to the low projected rotational velocity 
($\lesssim 2$~$\mathrm{km}\,\mathrm{s}^{1}$) and the H$_\alpha$ (6562.808 \AA)  
 line observed in absorption, that are signatures of  a  mature star older than
 $\sim$300 Myr.  
 The same authors suggest that GJ3470 is younger than 3 Gyr, since its
 kinematic properties match those of the young disk  population.
 Indeed, Fukui et al. 2013 performed a photometric monitoring of
 GJ3470, finding a peak-to-valley variability in the $\ic$ band of
 the order of 1\% in an interval of $\sim$60 days.  Assuming that
 such a variability is due to the presence of starspots, these have
 a filling factor $f$ of $\sim$ 0.01  \citep{ballerini2012}.  In
 Eq.~\ref{ratioff}, the effects of the unocculted and occulted spots
 tend to compensate each other, although the filling factor of the
 occulted spots may generally be higher because starspots tend to
 appear at low or intermediate latitudes in Sun-like stars.  The
 filling factor of the out-of-transit spots $\fu$ can be assumed to
 be constant because the visibility of those spots is modulated on
 timescales much longer than that of the transit, i.e., of the order
 of the stellar rotation period, which is generally several
 days. Instead, the filling factor of the occulted spots $\fo$ is in
 general a function of the position of the center of the planetary
 disc along the transit chord. A variable spot distribution in the
 occulted area gives rise to bumps in the observed transit light
 curve, whereas  the lack of obvious bumps in our data
 (Fig.~\ref{lcs}) suggests a more or less
 continuous background of spots along the transit chord. In the
 observed case, $\fo$ can be assumed to be almost constant during
 the transit because otherwise we would have detected the individual
 spot bumps. 

 The maximum deviation in the relative radius $k$ due to starspots
 can be computed in the unphysical hypothesis that all the spots are
 concentrated in the occulted area ($\fo=0.01$ and $\fu=0$) or outside
 the occulted area ($\fo=0$ and $\fu=0.01$).  Using the grid given by
 Ballerini et al. (2012), being GJ3470 very close to their case \#4,
 the maximum correction of the relative radius in our passbands 
 of interest is
 \begin{equation}
 \label{deltak}
 \begin{array}{lcl}
 \Delta k_{\max} \,(\uspec)  & = & 0.00540 \\
 \Delta k_{\max} \,(g')      & = & 0.00538 \\
 \Delta k_{\max} \,(\rc)     & = & 0.00529 \\
 \Delta k_{\max} \,(\ic)     & = & 0.00500 \\
 \Delta k_{\max} \,(F972N20) & = & 0.00500 \\
 \Delta k_{\max} \,(J)       & = & 0.00391 \\
 \Delta k_{\max} \,(4.5\mu\mathrm{m})  & = & 0.00355 \\
 \end{array}
 \end{equation}
These corrections should be considered only 
an upper limit to the actual impact of activity on our
estimated $k(\lambda)$. For this reason, in the 
following discussion and plots we will adopt the uncorrected values
of $k(\lambda)$ and $\rplanet(\lambda)$, unless noted
otherwise, and discuss separately the limit case
when  $\Delta k_{\max}$ is applied. 

A $\Delta k_{\max}$ correction ranging from 
0.00355 to 0.00540 could appear quite large since it is
an order of magnitude larger than our best errors on $k$
($\sim 0.0005$ in $F972N20$ and  $4.5\mu$m). 
One has to consider, however, that the \emph{absolute} scaling 
of this correction (which is related to the ``solid'' radius
of the planet $\rplanetzero$ measured at $H=0$) is of 
secondary interest for our study. Instead, the
\emph{differential}  effect of the $\Delta k_{\max}$ correction is
crucial, because it changes the slope of the Rayleigh
scattering absorption and the shape of the overall 
transmission spectrum. It is worth noting that
this fact may not hold when gathering transits at 
different epochs. In such a case, due to starspot 
evolution or activity cycles, the filling factors $\fo$ 
and $\fu$ may change, adding an unknown
systematic offset between non-simultaneous measurements. 
However, our main result is based on simultaneous measurements 
(in $\uspec$ and $F972N20$), 
hence they are not affected by such an effect.

\section{Discussion and conclusions}\label{discussion}

\subsection{The transmission spectrum of GJ3470b}

If we examine the transmission spectrum of GJ3470b as reconstructed by
our LBC measurements and from those by \citet{demory2013} and
\citet{fukui2013} (Fig.~\ref{spectrum}, large circles), it appears
evident  that no constant value of $k$ could explain the data
($\chi^2=67.8$ for 5 degrees of freedom, corresponding to a reduced
value of $\chi^2_\mathrm{r}=13.5$). The color dependence of $k$ is
then significant, with a steep increase toward the blue side of the
spectrum which is usually associated with scattering processes. How
can we interpret this spectrum $k(\lambda)$?  A detailed atmospheric
modeling of GJ3470b  is beyond the scope of this paper.  Instead, we
rescaled the model transmission spectra computed  by \citet{howe2012}
taking the grid point of their simulations closer to GJ3470b ($\teq =
700 K$, $\mplanet = 10 \mearth$) and investigating which class of
atmospheric composition best matches the observed data points, and
which ones can be excluded.

The most striking result is that the large observed variations  of
$k(\lambda)$ are totally incompatible with all model sets  by
\citet{howe2012} having an atmosphere of high mean molecular weight
$\mu$,  such as those made of pure $\htwoo$ (green line in
Fig.~\ref{spectrum}, upper panel) or pure methane ($\chfour$). All
those models predict a very flat spectrum due to their tiny scale
height, and are rejected by the available  data at $> 10\sigma$. The
same conclusion can be drawn by  considering the LBC data alone, as
$k(\uspec)$ and $k(F972N20)$ differ by $6\sigma$.

On the other hand, we investigated the fitness of   cloud-free,
haze-free atmospheric models of low $\mu$,  that is largely dominated
by H and He and with a large scale height $H$. This is the most
probable atmospheric composition expected for GJ3470b following
current  theories on planetary interiors
\citep{demory2013,rogers2010}.  A metal-poor model atmosphere having a
composition  with $Z=0.3\times Z_\odot$ rescaled from a solar
abundance (\citealt{howe2012}; orange line in the second panel of
Fig.~\ref{spectrum}) provides us with a much better fit on the red/IR
passbands (F972N20, $J$, $4.5\mu$m). Nevertheless  it fails at
reproducing the steep rise observed on the blue side of the spectrum,
especially for our $\uspec$ measurement.  That is, the scale height
required to output a Rayleigh  scattering from $\htwo$ large enough to
match our UV measurement would also give rise to infrared molecular
features which are inconsistent with the reddest data points, in
particular with the $4.5\mu$m one which is in a spectral region
particular  sensitive to species such as $\htwoo$ and CO.  Similar
models but with the addition of clouds perform even worse, flattening
the spectrum at intermediate wavelengths without significantly
improving the overall quality of the fit.

An additional source of scattering, other than $\htwo$, seems to
better explain the data. \citet{fortney2005} showed that  
a large variety of condensate
particles can be very efficient in this regard. 
Among them Tholin, for instance, is a 
polymer produced by UV  photochemistry of $\chfour$, suggested to
occur on  Jupiter and giant exoplanets \citep{sudarsky2003}.
Indeed, the
\citet{howe2012} class of models with a metal-poor H/He  dominated
atmosphere and high-altitude hazes from  Tholin
provide us with a satisfactory fit with $\chisqr = 5.2$ (black line in
Fig.~\ref{spectrum}, third panel; here the model points integrated
over the instrumental passbands  are plotted in magenta  for
clarity). Even after the maximum correction for unocculted spots
(Eq.~\ref{deltak}) has been  applied on all $k(\lambda)$ points, still
the  ``haze'' model with $\chisqr=7.9$ fits the data set better than a
flat function or any other class of  \citet{howe2012} models.
We emphasize that Tholin was chosen here only as a test model
to check whether the transmission spectrum of GJ3470b can
be explained through an additional source of scattering due to condensate particles. 
Still, we should keep in mind that a large number of molecules is able
to reproduce a similar result by means of the same physical
process, and their characteristic spectral fingerprints are so
subtle that they cannot
be discerned in a low-resolution spectrum. 
Additional high S/N observations will be required to confirm this
scenario, especially in the visible region at 400-900 nm (where the
S/N of the \citealt{fukui2013}  measurements is too low to constrain
the slope of the Rayleigh scattering independently from our data) and
in the near IR, for instance in the $H$ and $K$ spectral region, where
molecular bands such as $\htwoo$, CO or $\chfour$ are expected to
dominate the trasmission  spectrum giving rise to strong features
(Fig.~\ref{spectrum}). At  the estimated equilibrium temperature of
GJ3470b, $\teq\sim 700$~K, narrow alkali metal lines such as NaI and
KI begin to appear but are still too weak to hope in a ground-based
detection, even with the  current top-class facilities. Nevertheless
they become very prominent at 1000 K (Fig.~6 in \citealt{howe2012}),
so searching for them could put a tight and independent  upper limit
to the atmospheric effective temperature of GJ3470b. 

The available data can be interpreted also through a 
simplified analytical approach, first developed by
\citet{lecavelier2008a} under the usual approximation of
a well-mixed and isothermal atmosphere in chemical equilibrium. 
Following this work, we assume a scaling law of index $\alpha$
for the planetary cross section $\sigma=\sigma_0(\lambda/\lambda_0)^\alpha$. 
Under reasonable assumptions, the
slope of the planetary radius as a function of wavelength $\lambda$ can be
expressed as a function of the scale height $H$ (Eq.~\ref{scaleh}) as 
$\mathrm{d}R_\mathrm{p}/\mathrm{d}\!\ln\lambda = \alpha H$
\citep{lecavelier2008a}. The expression for 
the planetary equilibrium temperature then reduces to
\begin{equation}
\alpha T_\mathrm{eq} = \frac{\mu g}{k}\frac{\mathrm{d}R_\mathrm{p}}
{\mathrm{d}\!\ln\lambda}\textrm { .}
\label{teq}\end{equation}
If the main physical process involved is Rayleigh scattering and we
neglect atomic/molecular absorption, $\alpha = -4$ and
the mean molecular weight can be estimated as
\begin{equation}
\mu = -4 k T_\mathrm{eq}\left ( g \frac{\mathrm{d}R_\mathrm{p}}
{\mathrm{d}\!\ln\lambda} \right )^{-1}\textrm { .}
\label{molweight}\end{equation}
We assume from \citet{demory2013} 
a planetary surface gravity $g = 5.75$ $\mathrm{m}\,\mathrm{s}^{-2}$ and
an equilibrium temperature $T_\mathrm{eq}=(1-A_\mathrm{b})^{0.25}(683\pm 27)$ K.
As simple limiting cases, we can adopt two distinct values 
for the Bond albedo of GJ3470b: $A_\mathrm{b}=0.3$, 
which matches the albedo measured for 
Solar System icy giants (Uranus and Neptune); and $A_\mathrm{b}=0$,
corresponding to a cloud-free, very dark surface as 
theoretically predicted for some classes of ``temperate'' Neptunian
planets by \citet{spiegel2010}, and inferred from observations of the
hot Neptune GJ436b \citep{cowan2011}. With these assumptions the temperature 
becomes $\teq = 624 \pm 25$ K ($A_\mathrm{b}=0.3$),
and $\teq = 683 \pm 27$ K ($A_\mathrm{b}=0$), respectively.

\begin{figure}
\centering
\includegraphics[width=\columnwidth]{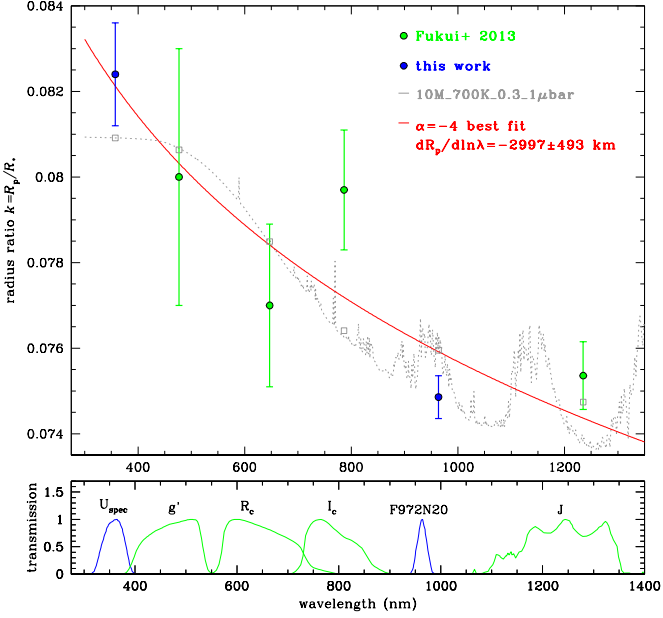}
\caption{\emph{Upper panel:} Same as Fig.~\ref{spectrum}, but
zoomed on the optical region. The red line corresponds to the best weighted 
linear fit in the $(\ln\lambda, k)$ plane over all data points 
at $\lambdac < 1 \mu$m (ours and from \citealt{fukui2013})
under the assumption of pure Rayleigh scattering; see text for details).
\emph{Lower panel:} Instrumental passbands employed
for each data point.
}
\label{slope}
\end{figure}

By fitting a straight line in the $(k,\ln\lambda)$  plane by weighted
least squares and estimating the uncertainties through an ordinary
bootstrap algorithm we find 
\begin{equation}
\begin{array}{ll}
k = (-0.0076\pm 0.0013) \ln\lambda + (0.123\pm 0.008) & \; (A_\mathrm{b}=0.3) \\
k = (-0.0075\pm 0.0012) \ln\lambda + (0.127\pm 0.008) & \; (A_\mathrm{b}=0)   \\
\end{array}
\end{equation}
which is equivalent to $\mathrm{d}R_\mathrm{p}/\mathrm{d}\!\ln\lambda
\simeq -3000 \pm 500$ km for both cases 
once we adopt $\rstar = 0.568$~$\rsun$ as the
stellar radius from \citet{demory2013} and propagate the uncertainties.  From Eq.~\ref{molweight}, and
taking into account the errors on all variables, we derive a mean
molecular weight of $\mu = 1.20 _{-0.17} ^{+0.24}$ atomic mass units (amu; $A_\mathrm{b}=0.3$) or
$\mu = 1.32 _{-0.19} ^{+0.27}$ amu ($A_\mathrm{b}=0$)
for GJ3470b,
that is a mean molecular weight consistent with that of  a
H/He-dominated atmosphere with sub-solar chemical abundances. The
error bars are largely dominated by the uncertainty on
$k$. Our
result is virtually independent on the correction $\Delta k _{\max}$
in Eq.~(\ref{deltak}), which leads to a  nearly unchanged $\mu =
1.3\pm 0.25$ amu ($A_\mathrm{b}=0.3$) when it is applied.
We emphasize however that the last approach does not take into account
that the $k$ value at $4.5\mu$m  seems to be inconsistent with a
cloud-free H/He envelope.  On the other hand, the $(k,\ln\lambda)$ fit
also demonstrate that the increase of $k$ (or $\rplanet$) at blue
wavelengths is  significant at $5.8\sigma$, whatever the underlying
physical process is.  The significance is largely  driven by our
$\uspec$ point, and this emphasize the importance of the optical UV
bands in exoatmospheric studies.

\begin{figure}
\centering
\includegraphics[width=\columnwidth]{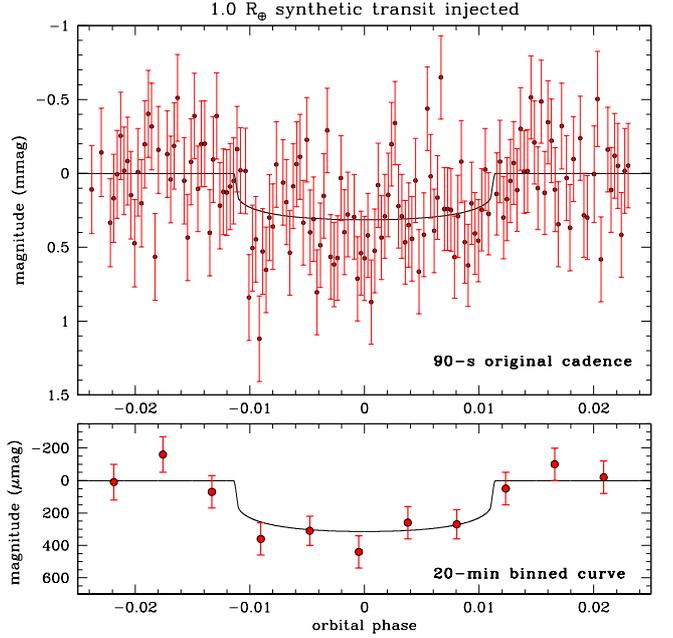}
\caption{\emph{Upper panel:} 
Synthetic 1-$R_\oplus$ transit (black line)
injected in the residuals of our $F972N20$ light curve (red points and error bars),
showing that a short-period Earth around a M1.5V star would have been detectable from 
our photometry. \emph{Lower panel:} same as above, binned over
20-min intervals.}
\label{earth}
\end{figure}

\subsection{Pushing the limits of ground-based photometry}

As a final note we highlight that the accuracy of our $F972N20$  light
curve is close to the theoretical limit expected for a photometric
series gathered with a 8.4-m telescope. It is interesting to
investigate the consequences of this unprecedented accuracy on the
ground-based detection of very small planets. A way to do it is to
extract the $F972N20$ residuals from our best-fit model (plotted as a
black line in Fig.~\ref{lcs}) and to inject on them a synthetic
transit corresponding to a $1$-$\rearth$ planet hosted by a star
having the same radius of  GJ3470, i.e.~a M1.5V dwarf. Having employed
real residuals to  perform this simulation, we are confident that the
instrumental,  atmospheric and  astrophysical noise contributions are
preserved.  The result is plotted in Fig.~\ref{earth}, with the
original 90-s sampling rate (upper panel) and binned over 20-minute
intervals to increase the S/N (lower panel). The transit signature is
clearly  detectable with high confidence ($9\sigma$),  opening new
prospects for a future targeted search for short-period, Earth-sized
transiting  planets around M dwarfs without necessarily resorting to
much  more expensive, dedicated space missions. 

The short-term photometric accuracy of the data presented here is
probably not achievable on a longer-term (night to night) baseline,
due to atmospheric and  instrumental systematic drifts. This
limitation, combined with the extremely low density of bright M dwarfs
in the sky prevents us from a  ``classical'' transit search of
terrestrial planets with a camera  such as LBC.  However, when a
sample of low-mass planets hosted by red dwarfs and discovered by a RV
survey will be available, it will be possible to  predict the instant
of the inferior conjunction within a few hours of uncertainty and to
search for planetary transits in that observational window. For a
scaled-down version of GJ3470b, the \emph{a priori} transit
probability is around $8\%$, meaning that at least a dozen of RV
planets need to be followed up in order to detect one transit. Given
the importance of such planets for the exoplanetary studies, thanks
especially to their closeness to the habitable zone and to the
relatively easy access to their atmospheric signatures, we suggest
that such a targeted search would be  worth to be implemented.

\begin{acknowledgements}

This work was partially supported by PRIN INAF 2008 ``Environmental
effects in the formation and evolution of extrasolar planetary
system''.  V.~N.~and G.~P.~acknowledge partial support by the
Universit\`a di Padova through the ``progetto di Ateneo \#CPDA103591''.
Some tasks of our data analysis have been carried out with
the VARTOOLS \citep{hartman2008} and \texttt{Astrometry.net} codes
\citep{lang2010}. This research has made use of the International 
Variable Star Index (VSX) database, operated at AAVSO, Cambridge, 
Massachusetts, USA.
\end{acknowledgements}

\bibliographystyle{aa}
\bibliography{biblio}

\end{document}